\documentclass[12pt]{article}
\usepackage[utf8]{inputenc}
\usepackage{indentfirst}
\usepackage{amsmath,amsthm,amssymb}
\usepackage{graphicx}
\usepackage{setspace}
\usepackage{hyperref}
\usepackage[all]{hypcap}
\usepackage[affil-sl,auth-sc]{authblk}
\title{The scalar product of XXZ spin chain revisited. Application to the ground 
state at $\Delta=-1/2$}
\author{A Garbali%
\thanks{Electronic address: garbali@lpthe.jussieu.fr}}
\affil{Laboratoire de Physique Th\'eorique et Hautes \'Energies, CNRS 
UMR 7589 and Universit\'e Pierre et Marie Curie (Paris 6), 4 place 
Jussieu, 75252 Paris cedex 05, France}
\date{}

\begin{document}

\maketitle

\begin{abstract}
For the scalar product $S_n$ of the XXZ $s=1/2$ spin chain we derive a new determinant 
expression which is symmetric in the Bethe roots. 
We consider an application of this formula to the inhomogeneous groundstate of the model 
with  $\Delta=-1/2$ with 
twisted periodic boundary conditions. At this point the ground state eigenvalue $\tau_n$ 
of the transfer matrix is known (see e.g.\cite{PDFPZJ}) and has a simple form that does not 
contain the Bethe roots. We use the knowledge of $\tau_n(\mu)$ to obtain a closed 
expression for the scalar product. The result is written in terms of Schur functions.
The computations of the normalization of the ground state and the expectation value of 
$\sigma^z$ are also presented.
\end{abstract}

\section{Introduction}
The computation of the correlation functions of the integrable XXZ spin-$1/2$ 
chain of finite length $N$ can be done using the form factor approach \cite{KBI,KMT}. 
These are the form factors of 
the local spin operators, they admit a nice determinant representation based on the 
Slavnov determinant \cite{S1,S2}. The Slavnov determinant is a scalar product of  
two states in the Algebraic Bethe Ansatz picture. Each of these states depend on 
the corresponding sets of rapidities of which one set is taken free and the other one is the 
set of Bethe roots. Thus, it is a scalar product between a Bethe state and an off-shell 
state. The Bethe roots satisfy a system of nonlinear algebraic equations (the Bethe equations). 
At this point, to obtain an explicit answer for the scalar products  
one must know the solution of the Bethe equations. In general, analytical computations 
for finite systems stop here. In the case when the interaction parameter $\Delta=-1/2$ 
(or $q^3=1$, where $q$ is the deformation parameter of the $U_q(A_1^{(1)})$ quantum 
group), called the combinatorial point of the model, it is possible to obtain a formula  
for the ground state scalar product which has no Bethe roots dependence. The answer is 
given in terms of a Schur function and its derivation is the purpose of the present work.

The twisted XXZ spin-$1/2$ chain at $\Delta=-1/2$ attracted a lot of attention in the past 
fifteen years due to the relation of its ground state with other statistical models. 
The numbers appeared in the ground state eigenvector of the 
transfer matrix are related to the combinatorics of the alternating sign matrices 
\cite{BGN,RS}. A spin chain state can be mapped to the so called loop basis, 
relating the XXZ to the dense Temperley-Lieb (TL) loop model.
In the TL model the condition $\Delta=-1/2$ is 
translated to the fact that the loop weight $n$ is equal to $1$. In this case this 
loop model describes interesting statistical systems like the critical bond percolation. 
A number of works \cite{MitraNienhuis,MitraNienhuis2,PDFBernard,dGNP,IP} were 
devoted to study correlation functions of this model. Due to the relation between the 
ground state of the spin chain and the ground state of the loop model the knowledge of 
the form factors on the spin side may serve as a tool for the computation of some 
correlation functions in the loop model.

The technical overview of our work is the following.
Our derivation of the ground state scalar product at $\Delta=-1/2$ and twist 
$\kappa=q^{2}$
is based on a few important aspects. First, we rewrite the scalar product in a 
symmetrized form. 
In the original determinant expression for the scalar product of an $n$-particle state: 
\begin{align}\label{sp0}
S_n\propto \det_{1\leq i,j\leq n}(s_{i,j}),
\end{align}
each matrix element $s_{i,j}$ is a function of the Bethe roots, which will be called 
$\zeta_1,..,\zeta_n$, a set of free parameters, called $\mu_1,..,\mu_n$ and a set of 
inhomogeneities $z_1,..,z_{N}$. The functions $s_{i,j}$ can be written as derivatives 
of the $n$-particle eigenvalue of the transfer matrix $\tau_n(\mu_i|\zeta_1,..,\zeta_n)$ 
with respect to the Bethe roots:
\begin{align}\label{spij}
s_{i,j}=\frac{\partial \tau_n(\mu_j|\zeta_1,..,\zeta_n)}{\partial{\zeta_{i}}}.
\end{align}
The functions $s_{i,j}$ are not symmetric in the Bethe roots 
in the original determinant expression \cite{S1,S2} and after taking the determinant 
we get extra factors of a Vandermonde determinant of $\zeta$'s and of $\mu$'s. If $s_{i,j}$ were symmetric 
in $\zeta$'s we could simply rewrite them
in terms of symmetric combinations of the Bethe roots, say, in terms of the elementary 
symmetric polynomials. Then, using the knowledge of the $Q$-function (computed in 
\cite{PDFPZJ}) for our special 
case, we could eliminate the elementary symmetric polynomials of the Bethe roots and obtain 
the desired formula. However, the functions $s_{i,j}$ are not symmetric in 
the Bethe roots. To overcome this, we use a symmetrization procedure to write $S_n$ as:
\begin{align}\label{sp1}
S_n\propto \det_{1\leq i,j\leq n}(\tilde{s}_{i,j}),
\end{align}
where now $\tilde{s}_{i,j}$ depend on the Bethe roots symmetrically. 
This is the same symmetrization which is used to show that the Jacobi-Trudi 
determinant of homogenious symmetric polynomials is equal to a Schur function \cite{MD}.

Next we rewrite 
$\tilde{s}_{i,j}$ in terms of the Baxter's $Q$-function and the $F$-function, that we 
define as follows: 
\begin{align}\label{FQ}
F_n(x)= \prod_{i=1}^{2n}(x-q^2 z_i).
\end{align}
Next, we use the T-Q relation \cite{Baxter} of the twisted XXZ spin chain. 
It allows to write the matrix 
elements $\tilde{s}_{i,j}$ back in terms of the eigenvalues $\tau_n$ without any 
derivatives. Note that $\tau_n$ is symmetric with respect to the Bethe roots.
Our expression for the scalar products in terms of $\tau_n$ (not the derivatives of 
$\tau_n$) resable the expression obtained in the framework of separation of 
variables \cite{Niccoli} and also expressions for the XXX spin$-1/2$ chain 
\cite{KM}. 
Until this point we assume generic $q$, generic twist $\kappa$ and also 
the numbers $N$ and $n$.

The second important aspect is the knowledge of the ground state eigenvalue 
(i.e. when $N=2n$) of the 
transfer matrix $\tau_n$ at $q^3=1$ and $\kappa=q^2$. This eigenvalue is given as a ratio 
of two $F$ functions which depend only on a paramter $\mu$ and the inhomogeneities $z$'s. 
Thus we rewrite the scalar product $S_n$ completely in terms of the $F$-functions. Expanding 
the $F$ functions in terms of the elementary symmetric polynomials we observe that our 
resulting matrix is, in fact, a product of two matrices. The scalar product $S_n$ becomes a
product of two determinants one of which is the Weyl formula for a Schur function of 
$\mu_1,..,\mu_{n}$ while the other one is the dual Jacobi-Trudi formula for a Schur function 
of $z_1,..,z_{2n}$.

The outline of the paper is the following. We give the introductory details on the 
Algebraic Bethe ansatz for the XXZ spin-$\frac{1}{2}$ model in the second section. 
Then, in the third 
section, we write in detail the first aspect described above. The second aspect of our 
derivation with the result for the scalar product are present in the fourth section. 
In the end of this section we include the computation of the norm of the ground state.
The fifth section contains the computation of the$\sigma^z_m$ expectation value.
After that we summarize the result of our work and discuss few open problems in 
the conclusion section.
The appendix explans how to go from symmetric polynomials of the Bethe roots 
to the symmetric polynomials of the rapidities.

\section{The XXZ Heisenberg spin-\texorpdfstring{$\frac{1}{2}$}{Lg} inhomogeneous finite chain}
In this chapter we will introduce the model and set the notations and conventions. The 
algebraic Bethe ansatz for this model can be found e.g. in \cite{Faddeev}. 
The XXZ Heisenberg model is defined by the following Hamiltonian:
\begin{align}\label{HH}
H_{XXZ}=J\sum_{i=1}^{N}\big{\{} \sigma_{i}^x\sigma_{i+1}^x+
\sigma_{i}^y\sigma_{i+1}^y +\Delta(\sigma_{i}^z\sigma_{i+1}^z-1)\big{\}},
\end{align}
where $\sigma_i^x,~\sigma_i^y$ and $\sigma_i^z$ are the standard Pauli matrices and the 
conditions $\sigma_{N+1}^x=\sigma_1^x$, $\sigma_{N+1}^y=\sigma_1^y$ and 
$\sigma_{N+1}^z=\sigma_1^z$ reflect the periodicity of the system. 
Now, we switch to
\begin{align}\label{sigma}
\sigma_i^{+}=(\sigma_i^x+i \sigma_i^y)/2,~~~\sigma_i^{-}=(\sigma_i^x-i \sigma_i^y)/2,
\end{align}
and the conditions $\sigma_{N+1}^{+}=e^{i\phi}\sigma_1^{+}$, 
$\sigma_{N+1}^-=e^{-i\phi}\sigma_1^-$ and 
$\sigma_{N+1}^z=\sigma_1^z$ reflect the twisted periodicity with the 
twist parameter $\phi$.

The Algebraic Bethe ansatz is formulated using the six vertex model. This model 
is defined by the $R$-matrix
\begin{align}\label{R}
R(\mu,t)=\left(
\begin{array}{cccc}
 a(\mu,t) & 0 & 0 & 0 \\
 0 & b(\mu,t) & c(\mu,t) & 0 \\
 0 & c(\mu,t) & b(\mu,t) & 0 \\
 0 & 0 & 0 & a(\mu,t ) \\
\end{array}
\right),
\end{align}
with the following weights
\begin{align}\label{wr}
&a(\mu,t)=\frac{q^2 \mu - t}{q(\mu - t)},~~~b(\mu,t)=1,\nonumber\\
&c(\mu,t)=\frac{(q^2-1)\sqrt{\mu~t}}{q(\mu - t)}. 
\end{align}
The $R$-matrix acts as a linear operator on two linear spaces $V_i$ and $V_j$ which are 
isomorphic to $\mathbb{C}^2$. To the spaces $V_i$ and $V_j$ we associate the spectral 
parameters $\mu_i$ and $\mu_j$, hence the $R$-matrix acting on $V_i\otimes V_j$ depends on 
$\mu_i$ and $\mu_j$ and is denoted by $R_{i,j}(\mu_i,\mu_j)$. Such an $R$-matrix 
satisfies the Yang-Baxter equation:
\begin{align}\label{YB}
R_{i,j}(\mu_i,\mu_j)R_{i,k}(\mu_i,\mu_k)R_{j,k}(\mu_j,\mu_k)=
R_{j,k}(\mu_j,\mu_k)R_{i,k}(\mu_i,\mu_k)R_{i,j}(\mu_i,\mu_j).
\end{align}
The two dimensional Hilbert space $\mathcal{H}_j$ of $SU(2)$ spin-$1/2$ chain corresponds 
to the $i$'th site of the chain. We construct the $L$ matrix acting on the site $j$: 
\begin{align}\label{L}
L_j(\mu,z_j)=R_{0,j}(\mu,z_j).
\end{align}
Thus the $L$ operator acts on the tensor product $\mathbb{C}^2\otimes\mathcal{H}_j$ and 
the parameter $z_j$ is an arbitrary complex number called the inhomogeneity (spectral 
paramter) associated to the space $\mathcal{H}_j$. 
Now we use the $L$ operators to construct the monodromy matrix:
\begin{align}\label{T}
T(\mu)=J L_{N}(\mu,z_N)..L_{1}(\mu,z_1).
\end{align}
The matrix $J$ is the diagonal matrix that includes the twist. The matrix (\ref{T}) is 
represented in the space with the label $0$, called the auxiliary space, 
as a two dimensional matrix:
\begin{align}\label{tm}
T(\mu)=\left(
\begin{array}{cc}
 A(\mu) & B(\mu) \\
 \kappa C(\mu) & \kappa D(\mu) \\
\end{array}
\right),
\end{align}
where the matrix elements $A,~B,~C$ and $D$ are linear operators acting on the Hilbert 
space $\mathcal{H}=\otimes_{j=1}^{N}\mathcal{H}_j$ of the chain of length $N$. 
The parameter $\kappa$ is the twist parameter which is related to the parameter $\phi$. The 
operators $A,~B,~C$ and $D$ form the Yang-Baxter algebra whose commutation relations 
follow from the RTT relation:
\begin{align}\label{RTT}
R_{1,2}(\mu_1,\mu_2)T_1(\mu_1)T_2(\mu_2)=T_2(\mu_2)T_1(\mu_1)R_{1,2}(\mu_1,\mu_2),
\end{align}
where we used the notation $T_1(\mu)=T(\mu)\otimes Id$ and $T_2(\mu)=Id\otimes T(\mu)$.

The trace of the monodromy matrix eq.(\ref{tm}) over the auxiliary space defines the 
transfer matrix $\mathcal{T}(\mu)=A(\mu)+\kappa D(\mu)$. Thanks to the Yang-Baxter 
equation the transfer matrices at different values of the spectral parameter $\mu$ commute.
The Hamiltonian $H_{XXZ}$ can be written in terms of the transfer matrix $\mathcal{T}(\mu)$ 
hence the problem turns into the diagonalization of $\mathcal{T}(\mu)$ for all values 
of $\mu$. 

In the algebraic Bethe ansatz the eigenvectors of the transfer matrix can be written 
in terms of the $B$-operators acting on the reference state $|0\rangle$. This state 
must have the following properties:
\begin{align}\label{vac}
&A(\mu)|0\rangle=a(\mu)|0\rangle,\nonumber\\
&D(\mu)|0\rangle=d(\mu)|0\rangle,\nonumber\\
&C(\mu)|0\rangle=0,\nonumber\\
&B(\mu)|0\rangle\neq 0.
\end{align}
The functions $a(\mu)$ and $d(\mu)$ are defined as:
\begin{align}\label{ad}
&a(\mu)=\prod_{i=1}^N a(\mu,z_i),~~~d(\mu)=\prod_{i=1}^N b(\mu,z_i).
\end{align}
The $|0\rangle$ state is totally ferromagnetic in the case of the XXZ model. The eigenvectors 
of the transfer matrix are obtained by the successive action of the $B$-operators on the 
reference state:
\begin{align}\label{Bstate}
\psi_n=\prod_{i=1}^n B(\zeta_i|z_1,..,z_N)|0\rangle,
\end{align}
where the parameters $\zeta_1,..,\zeta_n$ satisfy the Bethe equations:
\begin{align}\label{Beq}
\prod_{i=1}^N a(\zeta_k,z_i)\prod_{\substack{i=1 \\ i\neq k }}^n 
\frac{a(\zeta_i,\zeta_k)}{c(\zeta_i,\zeta_k)}-(-1)^n \kappa 
\prod_{i=1}^N b(\zeta_k,z_i)\prod_{\substack{i=1 \\ i\neq k }}^n 
\frac{a(\zeta_k,\zeta_i)}{c(\zeta_k,\zeta_i)}=0,~~~~k=1,2,..n.
\end{align}
Taking into account (\ref{wr}) we may write these equations simply:
\begin{align}\label{Beq2}
\prod_{i=1}^N a(\zeta_k,z_i)\prod_{\substack{i=1 \\ i\neq k }}^n 
a(\zeta_i,\zeta_k)- \kappa 
\prod_{i=1}^N b(\zeta_k,z_i)\prod_{\substack{i=1 \\ i\neq k }}^n 
a(\zeta_k,\zeta_i)=0,~~~~k=1,2,..n.
\end{align}
A state (\ref{Bstate}) with the parameters $\zeta_1,..,\zeta_n$ that satisfy (\ref{Beq})
is called the $n$-particle Bethe state and the parameters $\zeta_1,..,\zeta_n$ are called the 
Bethe roots. We will reserve the indexed letter $\zeta$ in what follows to denote the 
Bethe roots, the parameters $z_1,..,z_N$ will denote the inhomogeneities of the system 
and the parameters $\mu_1,..,\mu_n$ will be another set of free parameters which is necessary 
in order to write the scalar products. The transfer matrix eigenvalue corresponding to 
the $n$-particle state is
\begin{align}\label{tev}
\tau_n(\mu)=\prod_{i=1}^N a(\mu,z_i)\prod_{i=1}^n a(\zeta_i,\mu)+
\kappa \prod_{i=1}^N b(\mu,z_i)\prod_{i=1}^n a(\mu,\zeta_i).
\end{align}

The scalar products of states are defined as 
\begin{align}\label{sp}
S_n(\mu_1,..,\mu_n;\zeta_1,..,\zeta_n)=
\langle 0|\prod_{i=1}^{n}C(\mu_i)\prod_{i=1}^{n}B(\zeta_i)|0\rangle.
\end{align}
Here, as mentioned before $\zeta_1,..,\zeta_n$ are the Bethe roots and 
$\mu_1,..,\mu_n$ are free parameters. If the parameters $\mu$ also satisfy the 
Bethe equations, then the product of the $C$ operators acting on the dual reference state 
\begin{align}\label{dualv}
\langle 0|\prod_{i=1}^{n}C(\mu_i)
\end{align}
is the dual Bethe state. If we want to compute the expectation value of an operator 
$\mathcal{O}$:
\begin{align}\label{evO}
\langle \mathcal{O} \rangle =
\langle 0|\prod_{i=1}^{n_0}C(\mu_i)\mathcal{O}\prod_{i=1}^{n}B(\zeta_i)|0\rangle,
\end{align}
and, say, we computed the action of $\mathcal{O}$ on the dual Bethe state written as 
a combination of dual states
\begin{align}\label{evO2}
\langle 0|\prod_{i=1}^{n_0}C(\mu_i)\mathcal{O}=\sum_{k}\theta_k 
\langle 0|\prod_{i=1}^{n_k}C(\nu_i^{(k)}),
\end{align}
where $\nu_i^{(k)}$ are some numbers, then the computation of 
$\langle \mathcal{O} \rangle$ boils down to the computation 
of the scalar product (\ref{sp}).

Fortunately, the scalar products (\ref{sp}) have a nice determinantal representation 
\cite{S1,S2}.
Let us introduce the matrix elements $\Omega_{j,k}$ which depend on the system length $N$ 
and the number of particles $n$: 
\begin{align}\label{omega}
\Omega_{j,k}=\frac{\partial \tau_n(\mu_k|\zeta_1,..,\zeta_n)}{\partial{\zeta_{j}}}
\prod_{i=1}^{n}c^{-1}(\zeta_i,\mu_k).
\end{align}
Then the scalar product $S_n$ is
\begin{align}\label{spSL}
S_n(\mu_1,..,\mu_n;\zeta_1,..,\zeta_n)=\frac{(q^2-1)^n}{2^n}
\prod_{i< j}c(\mu_j,\mu_i)c(\zeta_i,\zeta_j)
\det_{1\leq j,k \leq n}\Omega_{j,k}.
\end{align}
Let us rewrite slightly this expression. We introduce two functions that we will use 
later on:
\begin{align}\label{F}
F_N(x)=\prod_{i=1}^N(x-q^2 z_i),
\end{align}
and 
\begin{align}\label{Q}
Q_n(x)=\prod_{i=1}^n(x-\zeta_i).
\end{align}
Where $Q_n(x)$ is nothing but the Baxter's $Q$-function \cite{Baxter} corresponding to 
the $n$-particle state. The roots of the $Q$-polynomial are the Bethe roots. 
We will often omit the indices in $F$ and $Q$, since usually this is clear from the context. 
In terms of $F$ and $Q$ the matrix elements $\Omega_{j,k}$ become:
\begin{align}\label{omega2}
\Omega_{j,k}=
\frac{-2(q^2-1)\zeta_j}{(1-q^2)^n F(q^2 \mu_k)\mu_k^{n/2-1}\prod_{i\neq j}\zeta_i^{1/2}}
\times c_{j,k},
\end{align}
where we gathered the nontrivial part of the matrix elements $\Omega_{j,k}$ into $c_{j,k}$:
\begin{align}\label{cjk}
c_{j,k}=\frac{1}{\mu_k-\zeta_j}\bigg{(} \frac{Q(q^{-2}\mu_k) F(q^{4}\mu_k)}{\mu_k-q^2\zeta_j}+ 
\kappa \frac{Q(q^{2}\mu_k) F(q^{2}\mu_k)}{\zeta_j-q^2\mu_k} \bigg{)}.
\end{align}
The prefactor of $c_{j,k}$ in (\ref{omega2}) can be taken out from the sign of the 
determinant in (\ref{spSL}), so now we need to compute: 
\begin{align}\label{detcjk}
\tilde{S}_n=\frac{1}{\prod_{i< j}(\zeta_j-\zeta_i)(\mu_j-\mu_i)}
\det_{1\leq j,k\leq n}c_{j,k}.
\end{align}

\section{Symmetric expression for the scalar products}
In this section we show the derivation of a symmetric expression for the matrix $c_{j,k}$. 
We first split the matrix 
$c_{j,k}$ into two parts proportional to $a_{j,k}$ and $b_{j,k}$:
\begin{align}
&a_{j,k}=\frac{1}{(\mu_k-\zeta_j)(\mu_k-q^2\zeta_j)},\label{amat}\\
&b_{j,k}=\frac{1}{(\mu_k-\zeta_j)(\zeta_j-q^2\mu_k)}\label{bmat},
\end{align}
so that 
\begin{align}\label{cjk2}
c_{j,k}=Q(q^{-2}\mu_k) F(q^{4}\mu_k) a_{j,k}+ 
\kappa Q(q^{2}\mu_k) F(q^{2}\mu_k) b_{j,k}.
\end{align}
Now we can symmetrize separately the $a_{j,k}$ and the $b_{j,k}$ parts. In order to do that 
we follow the procedure that allows to show, for example, that a Schur function can be 
written as a determinant of the homogenious symmetric functions \cite{MD}. 
First, define the operators $\mathcal{R}_{i}$ 
which act on a $n\times n$ matrix $A$ as follows:
\begin{align}\label{R1}
(\mathcal{R}_k A)_{i,j}= A_{i,j}-\delta_{i,k}A_{i+1,j}.
\end{align}
This operator subtracts two rows $i$ and $i+1$ which have the same form 
except that one depends on $\zeta_i$ and the other one depends on $\zeta_{i+1}$. The 
result of such subtraction is proportional to $\zeta_i-\zeta_{i+1}$ with the proportionality 
factor being symmetric in the interchange of $\zeta_i$ and $\zeta_{i+1}$. Let $A$ have 
components either $a_{i,j}$ or $b_{i,j}$, then we apply to it the product of the 
$\mathcal{R}_i$ operators:
\begin{align}\label{R1A}
&\mathcal{R}_{n-1}..\mathcal{R}_1 A = A^{(1)}\prod_{i=1}^{n-1}(\zeta_i-\zeta_{i+1}),
\nonumber\\
&\mathcal{R}_{n-2}..\mathcal{R}_1 A^{(1)} = A^{(2)}\prod_{i=1}^{n-2}
(\zeta_i-\zeta_{i+2}),\nonumber\\
&\vdots \nonumber \\
&\mathcal{R}_1 A^{(n-1)} = A^{(n)}(\zeta_1-\zeta_{n}).
\end{align}
The components of the resulting matrix $A^{(n)}$, apart from the first row, are not yet 
symmetric in the Bethe roots. 
On the way to $A^{(n)}$ we lost exactly the Vandermonde determinant of the Bethe roots as 
we wanted. The matrix elements of $A^{(n)}$ (recall that those are either $a_{j,k}$ or $b_{j,k}$) 
significantly differ in their form from one row to another since we made a different number of operations 
on different rows. We need to apply another transformation to the 
matrix $A^{(n)}$. Define the operators $\mathcal{R}_{i}(x)$ which act on a $n\times n$ 
matrix in the following way
\begin{align}\label{R2}
(\mathcal{R}_k(x) A)_{i,j}= A_{i,j}+\delta_{i,k}x A_{i-1,j}.
\end{align}
And now we do something similar as before
\begin{align}\label{R2A}
&\mathcal{R}_2(\zeta_1)..\mathcal{R}_{n}(\zeta_{n-1})A^{(n)} = A^{(n,1)},
\nonumber\\
&\mathcal{R}_3(\zeta_1)..\mathcal{R}_{n}(\zeta_{n-2})A^{(n,1)} = A^{(n,2)},
\nonumber\\
&\vdots \nonumber \\
&\mathcal{R}_{n}(\zeta_{1}) A^{(n,n-1)} = A^{(n,n)}.
\end{align}
The matrix $A^{(n,n)}$ is symmetric in the Bethe roots as we wanted. To see how do the matrix 
elements of $a$ and $b$ look we need to express the transformations of the first step 
(\ref{R1A}) and the transformations of the second step (\ref{R2A}) in terms of matrices. 
An operator $\mathcal{R}_i$ can be viewed as a simple matrix, hence the transformation, 
for example, of the first line of (\ref{R1A}) will be a product of such matrices divided 
by the prefactor of the $A^{(1)}$ matrix. The transformation of the first step 
(\ref{R1A}) will be the product of the matrices corresponding to the transformations 
of each line in (\ref{R1A}). Similar logic applies to the second step (\ref{R2A}). 
We denote these matrices by $\rho_1$ and $\rho_2$ respectively:
\begin{align}\label{rho1}
&\rho_{1}=
\left(
\begin{array}{ccccc}
 \frac{1}{\left(\zeta _1-\zeta _2\right) \left(\zeta _1-\zeta _3\right)\dots \left(\zeta
   _1-\zeta _n\right)} & \frac{1}{\left(\zeta _2-\zeta _1\right) \left(\zeta _2-\zeta
   _3\right)\dots \left(\zeta _2-\zeta _n\right)} & \dots & \frac{1}{\left(\zeta
   _n-\zeta _1\right) \left(\zeta _n-\zeta _2\right) \left(\zeta _n-\zeta _{n-1}\right)} \\
 0 & \frac{1}{\left(\zeta _2-\zeta _3\right)\dots \left(\zeta _2-\zeta _n\right)} &
   \dots &
   \frac{1}{\left(\zeta _n-\zeta _2\right)\dots \left(\zeta _n-\zeta _{n-1}\right)} \\
 \vdots & \vdots & \vdots & \vdots \\
 0 & 0 & \dots & \frac{1}{\zeta _n-\zeta _{n-1}} \\
 0 & 0 & 0 & 1 \\
\end{array}
\right)
\end{align}
\begin{align}\label{rho2}
\rho_{2}=
\left(
\begin{array}{ccccc}
1 & 0 & 0 & \dots & 0 \\
h_1(\zeta_1) & 1 & 0 & \dots & 0 \\
h_2(\zeta_1) & h_1(\zeta_1,\zeta_2) & 1 & \dots & 0 \\
\vdots & \vdots & \vdots & \dots & \vdots  \\
h_{n-1}(\zeta_1) & h_{n-2}(\zeta_1,\zeta_2) & h_{n-3}(\zeta_1,\zeta_2,\zeta_3) & \dots & 1 \\
\end{array}
\right),
\end{align}
where $h_{k}$ are the homogeneous symmetric functions:
\begin{align}\label{hsp}
h_{k}(x_1,..,x_n)=
\sum_{1\leq i_1 \leq i_2 \leq\dots \leq i_k \leq n}x_{i_1}x_{i_2}\dots x_{i_k},
\end{align}
which have, in particular, the following useful property: 
\begin{align}\label{hspprop}
h_{i-1}(x_1,..,x_m)=\frac{h_{i}(x_1,..,\hat{x}_l,..,x_m)-
h_{i}(x_1,..,\hat{x}_k,..,x_m)}{x_k-x_l},
\end{align}
where $\hat{x}$ means the absence of the corresponding variable.
Multiplying the two transformations $\rho_1$ and $\rho_2$ and using the property 
(\ref{hspprop}) we obtain the full transformation $\rho=\rho_2 \rho_1$:
\begin{align}\label{rho}
\rho_{i,j}=\frac{\zeta_j^{i-1}}{\prod_{k\neq j}(\zeta_j-\zeta_k)}.
\end{align}
This is a linear transformation, so we act separately on $a_{i,j}$ and $b_{i,j}$ and 
then add up the results with the appropriate coefficients according to (\ref{cjk2}). 
Let us take first the matrix $a_{i,j}$ and multiply it by the matrix $\rho$ from the left. 
We obtain:
\begin{align}\label{rhoa}
\rho_{i,j}a_{j,k}=\sum_{j=1}^n\frac{\zeta_j^{i-1}}{\prod_{l\neq j}(\zeta_j-\zeta_l)}
\times \frac{1}{(\mu_k-\zeta_j)(\mu_k-q^2\zeta_j)}.
\end{align}
Let us multiply ($\ref{rhoa}$) by the product $\mu_k^{2-i}Q(\mu_k)Q(q^{-2}\mu_k)$, we get
\begin{align}\label{rhoa2}
\sum_{j=1}^n\frac{\zeta_j^{i-1}}{\prod_{l\neq j}(\zeta_j-\zeta_l)}
\times \frac{\mu_k^{2-i}Q(\mu_k)Q(q^{-2}\mu_k)}{(\mu_k-\zeta_j)(\mu_k-q^2\zeta_j)}=
\sum_{j=1}^n q^{-2}\zeta_j^{i-1}\mu_k^{2-i} \prod_{l\neq j}\frac{(\mu_k-\zeta_l)
(\mu_k q^{-2}-\zeta_l)}{(\zeta_j-\zeta_l)}.
\end{align}
The right hand side here is the Lagrange polynomial of a function, let us call it 
$f(\mu_k)$. Now we set $\mu_k=\zeta_i$ for $i=1,..,n$. From (\ref{rhoa2}) we obtain
\begin{align}\label{f1}
f(\zeta_i)=\frac{1}{1-q^2}Q( q^{-2}\zeta_1).
\end{align}
If we set $\mu_k=q^{2}\zeta_i$ for $i=1,..,n$, then
\begin{align}\label{f2}
f(q^2\zeta_i)=q^{2-2i}\frac{1}{q^2-1}Q(q^{2}\zeta_1).
\end{align}
It is easy to check that
\begin{align}\label{fQ}
f(\mu)=\frac{Q(\mu)q^{2-2i}-Q(q^{-2}\mu)}{q^2-1},
\end{align}
satisfies the above $2n$ recurrence relations (\ref{f1}) and (\ref{f2}) being a polynomial 
of degree $n$, hence the Lagrange polynomial in (\ref{rhoa2}) is equal to 
the function (\ref{fQ}). The matrix elements $a_{i,j}$ after the transformation take the form:
\begin{align}\label{aQ}
\rho_{i,j}a_{j,k}=\mu_k^{i-2}\frac{Q(\mu_k)q^{2-2i}-Q(q^{-2}\mu_k)}
{(q^2-1)Q(\mu_k)Q(q^{-2}\mu_k)}.
\end{align}
Now we do the same with the $b_{i,j}$ matrix. The result will be
\begin{align}\label{bQ}
\rho_{i,j}b_{j,k}=\mu_k^{i-2}\frac{Q(\mu_k)q^{2i-2}-Q(q^{2}\mu_k)}
{(q^2-1)Q(\mu_k)Q(q^{2}\mu_k)}.
\end{align}
Finally, we can write the action of the transformation $\rho$ on the matrix elements 
$c_{i,j}$. We will call $\tilde{c}_{i,j}$ the matrix elements of the transformed 
Slavnov matrix $\tilde{c}_{i,k}=\rho_{i,j}c_{j,k}$  
\begin{align}\label{cQ}
\tilde{c}_{i,k}=\frac{\mu_k^{i-2}}{(q^2-1)}
\bigg{(} \frac{F(q^4\mu_k)}{Q(\mu_k)}(q^{2-2i} Q(\mu_k)-Q(q^{-2}\mu_k))-
\kappa \frac{F(q^2\mu_k)}{Q(\mu_k)}(Q(q^2\mu_k)-q^{2i-2}Q(\mu_k)) \bigg{)}. 
\end{align}
This expression already satisfies our needs. It is symmetric in the Bethe roots, and 
even more than that. We do not need to express it as elementary symmetric polynomials of the 
Bethe roots and then use the $Q$ function, since it is already written in terms of the $Q$ 
function. The determinant 
\begin{align}\label{SPQ}
\tilde{S}_n=\prod_{i< j}(\mu_i-\mu_j)^{-1}\det_{1\leq i,j\leq n} \tilde{c}_{i,j}
\end{align}
is written in a nice form, since it depends only on the $Q$ operators 
and a simple factorized function $F$. Similar formulae for the scalar product 
appear in the separation of variables approach, see \cite{Niccoli}.

Now let us make (\ref{cQ}) even nicer. For that we need to use the $T-Q$ relation for the 
six vertex model \cite{Baxter}. In this relation the transfer matrix eigenvalues and the 
$Q$ operator satisfy a quadratic relation, in our notations it reads
\begin{align}\label{TQ}
\tau(\mu)Q(\mu)=q^{-n}\frac{F(q^4\mu)}{F(q^2\mu)}Q(q^{-2}\mu)+q^{-n}\kappa Q(q^2\mu).
\end{align}
We can use this expression to simplify the matrix elements $\tilde{c}_{i,j}$. After this 
simplifications the $Q$-functions disappear. Indeed, the sum of the two
terms in (\ref{cQ}) which contain $Q(q^{-2}\mu)$ and $Q(q^{2}\mu)$ is proportional to 
$\tau(\mu)Q(\mu)$, the denominator of $Q(\mu)$ cancels in every term in this expression, 
therefore the remaining depends on the transfer matrix eigenvalue $\tau(\mu)$ and the 
polynomials $F$:
\begin{align}\label{cT}
\tilde{c}_{i,k}=\frac{\mu_k^{i-2}}{(q^2-1)}
\bigg{(} q^{2-2i} F(q^4\mu_k) +F(q^2\mu_k) (q^{-2+2i}\kappa -q^n \tau(\mu_k)) \bigg{)},
\end{align}
and recalling that the prefactor in (\ref{omega2}) contains $1/F(q^2 \mu_k)$, we write:
\begin{align}\label{cT2}
\bar{c}_{i,k}=\frac{\mu_k^{i-2}}{(q^2-1)}
\bigg{(} q^{2-2i} \frac{F(q^4\mu_k)}{F(q^2\mu_k)} +q^{-2+2i}\kappa -q^n \tau(\mu_k) \bigg{)},
\end{align}
Finally, the Slavnov product reads:
\begin{align}\label{spSL3}
\tilde{S}_n=\frac{1}{\prod_{i< j}(\mu_j-\mu_i)}
\det_{1\leq i,k \leq n}\mu_k^{i-2}
\bigg{(} q^{2-2i} \frac{F(q^4\mu_k)}{F(q^2\mu_k)} +q^{-2+2i}\kappa -q^n \tau(\mu_k) \bigg{)}.
\end{align}
This expression involves 
the eigenvalue of the transfer matrix which contains all the dependence on the Bethe roots 
and is symmetric in their interchange. When the first set of variables in the scalar 
product are the Bethe roots ($\mu_1,..,\mu_n$ in our notation) a similar formula to the 
(\ref{spSL3}) can be obtained.

\section{Scalar product at \texorpdfstring{$q^3=1$}{Lg}}
In this section we will set $q^3=1$, $\kappa=q^2$ and consider the scalar product in 
which $N=2n$, i.e. the one for the ground state vector. In this special case 
we know the eigenvalue of the transfer matrix. We will substitute it in the matrix 
elements $\tilde{c}_{i,j}$ in the form (\ref{cT}) and after some 
simplifications obtain a product of two Schur functions in which one Schur function 
depends on the inhomogeneities while the other one on the free parameters $\mu$. 

The ground state eigenvalue of the transfer matrix reads \cite{PDFPZJ}:
\begin{align}\label{Tev}
\tau(\mu)=-q^{2n+1}\frac{F(\mu)}{F(q^2\mu)}.
\end{align}
Matrix elements $\tilde{c}_{i,k}$ become
\begin{align}\label{ctau}
\tilde{c}_{j,k}=\frac{\mu_k^{j-2}}{q^2-1}
\big{(}q^{2-2j} F(q^4 \mu_k)+ q^{2j}F(q^2 \mu_k) +q F(\mu_k) \big{)}.
\end{align}
The functions $F$ are the generating functions of the elementary symmetric polynomials:
\begin{align}
&F_n(x)=\sum_{i=0}^{2n} (-q^2)^{2n-i} x^i e_{2n-i}(z_1,..,z_{2n}),\label{Fesp}\\
&e_k(z_1,..,z_m)=
\sum_{1\leq i_1 < i_2 <\dots < i_k \leq m}z_{i_1}z_{i_2}\dots z_{i_k},\\
&e_k(z_1,..,z_m)=0,~~\text{for $k<0$ or $k>m$}.\label{esp0}
\end{align}
Substituting (\ref{Fesp}) in (\ref{ctau}) we obtain: 
\begin{align}
\tilde{c}_{j,k}=\frac{\mu_k^{j-2}q^{4n}}{q^2-1}
\sum_{s=0}^{2n} (-1)^{s}\mu_k^s e_{2n-s}\big{(} q^{1-2s}+q^{2j}+q^{2+2s+2j}\big{)}, \nonumber
\end{align}
which can be rewritten as
\begin{align}\label{cesp}
\tilde{c}_{j,k}=\frac{\mu_k^{j-2}q^{n-j}}{q^2-1}
\sum_{s=0}^{2n} (-1)^{s}\mu_k^s e_{2n-s}\big{(} q^{1+s+j}+q^{-(1+s+j)}+1\big{)},
\end{align}
where we used $q^2=q^{-1}$ and also assumed $e_i=e_i(z_1,..,z_{2n})$. The factor 
$q^{1+s+j}+q^{-(1+s+j)}+1$ in the last expression is equal to zero unless $j+s+1=3m$, 
with $m\in \mathbb{Z}$, in which case it is equal to $3$. Noticing this, we change the 
summation $s=3m-j-1$: 
\begin{align}\label{cesp2}
\tilde{c}_{j,k}=\frac{3q^{n-j}}{q^2-1}
\sum_{m=\lceil \frac{j+1}{3}\rceil}^{\lfloor \frac{2n+j+1}{3}\rfloor} 
(-1)^{j+m+1}\mu_k^{3m-3} e_{2n-3m+j+1}.
\end{align}
Because of the property (\ref{esp0}) we can put the initial value of the summation in 
$m$ to $1$. For the first and second row ($j=1,2$) this is already true, but for the 
remaining rows the summation will start at a value $m>1$ and due to 
(\ref{esp0}) we can extend it to $m=1$. When the value of $m$ becomes higher than $n$ 
all terms vanish by the same property (\ref{esp0}), so we can simply put the 
upper summation limit to be $n$. Hence we can write:
\begin{align}\label{cesp3}
\tilde{c}_{j,k}=\frac{3q^{n-j}}{q^2-1}
\sum_{m=1}^{n} 
(-1)^{j+m+1}\mu_k^{3m-3} e_{2n-3m+j+1}.
\end{align}
The last expression is nothing but the product of two $n\times n$ matrices
\begin{align}\label{AB}
A_{k,m}= \mu_{k}^{3m-3},~~~B_{m,j}= e_{2n-3m+j+1},
\end{align}
multiplied by a prefactor that we can take out of the determinant. 
The matrix $A$ is the Schur polynomial of the partition $Y_n=\{2n-2i\}_{i=1}^{n}$ 
by the Weyl formula for the $GL(N)$ characters:
\begin{align}\label{Aschur}
\det A=s_{Y_n}(\mu_1,..,\mu_{n}).
\end{align}
This Schur function has a simple factorized form:
\begin{align}\label{Aschur2}
s_{Y_n}(\mu_1,..,\mu_{n})=\prod_{1\leq i<j \leq n}(\mu_i^2+\mu_i \mu_j+\mu_j^2).
\end{align}
The matrix $B$ is the Schur polynomial of the partition 
$\tilde{Y}_n=\{n,n-1,n-1,..,1,1,0\}$ by the dual Jacobi-Trudi identity
\begin{align}\label{Bschur}
\det B=s_{\tilde{Y}_n}(z_1,..,z_{2n}).
\end{align}

The determinant $S_n$ becomes:
\begin{align}\label{spscur}
S_n=\frac{3^n q^n\prod_{i=1}^{n}\mu_i^{1/2}\zeta_i^{1/2}}{\prod_{i=1}^{n}q^{-4 n}F(q^2\mu_i)} 
s_{Y_n}(\mu_1,..,\mu_n) s_{\tilde{Y}_{n}}(z_1,..,z_{2n}).
\end{align}
Because of the simplicity of this expression we hope that it will be 
helpful in the problem of the computation 
of correlation functions. 

Before turning to the problem of the computation of the  
form factor $\langle \sigma^z \rangle$ we derive the normalization of the ground state 
$\mathcal{N}_n$. It was computed previously in \cite{Cant} and also in \cite{Paul}.
This quantity is obtained from the eq.(\ref{spscur}) 
by assuming $\mu_1,..,\mu_n$ to be the Bethe roots. In the appendix we will show 
that a Schur function $s_{\pi}(\zeta_1,..,\zeta_n)$ of a partition $\pi$, which is a symmetric 
polynomial of the Bethe roots $\zeta_i$, is equal to some polynomial 
$p_{\pi}(z_1,..,z_{2n})$ which is labeled by the same partition $\pi$, but depends on 
the inhomogeneities. It turns out, however, that in the particular case of the Schur function 
of the partition $Y_n$, which appears in (\ref{spscur}),
we can write $s_{Y_n}$ in a more explicit form. Setting $\mu_i=\zeta_i$ in 
(\ref{spscur}) we get:
\begin{align}\label{norm1}
\mathcal{N}_n=\frac{3^n q^n\prod_{i=1}^{n}\zeta_i}{\prod_{i=1}^{2n}Q(z_i)} 
s_{Y_n}(\zeta_1,..,\zeta_n)s_{\tilde{Y}_{n}}(z_1,..,z_{2n}).
\end{align}
Now we need to know what is $s_{Y_n}(\zeta_1,..,\zeta_n)\prod_{i=1}^{n}\zeta_i$. Taking into 
accound (\ref{Aschur2}) and that $q^3=1$ we can rewrite this as: 
\begin{align}\label{norm2}
s_{Y_n}(\zeta_1,..,\zeta_n)\prod_{i=1}^{n}\zeta_i=(q-1)^{-n}\prod_{i=1}^nQ(q \zeta_i).
\end{align}
Now we use the identity
\begin{align}\label{Qprod}
\prod_{i=1}^nQ(q \zeta_i)=q^{2n^2-n}(1-q)^n\frac{s_{\tilde{Y}_n}(z_1,..,z_{2n})
s^2_{Y^{\prime}_n}(z_1,..,z_{2n})}{s_{Y^0_n}(z_1,..,z_{2n})}
\prod_{i=1}^{2n}\frac{Q(z_i)}{F(z_i)},
\end{align}
where the partition $Y_0$ is simply $Y_0=\{1,1,..,1\}$ and the partition $Y_{2n}^{\prime}$ 
is another staircase partition: $Y_{2n}^{\prime}=\{n,n,..,1,1\}$. 
The entries of the equation (\ref{Qprod}) satisfy certain recurrence relations upon setting 
$z_i=q z_j$. The explicit form of $Q$ in terms of inhomogeneities 
is known \cite{PDFPZJ}, and we remind it in the appendix (\ref{Qesp}). These two ingredients 
are enough to prove (\ref{Qprod}). It would be, however, much more usefull to have a direct proof 
of the eq.(\ref{Qprod}).

The product of the polynomials $F$ in the denominator in (\ref{Qprod}) 
can be expressed through the the product of Schur function of partitions 
$Y_{2n}$ and of $Y^0_n$, hence, omitting the irrelevant prefactor depending on $q$, we get:
\begin{align}\label{norm3}
\mathcal{N}_n=\frac{s^2_{\tilde{Y}_n}(z_1,..,z_{2n})
s^2_{Y^{\prime}_{2n}}(z_1,..,z_{2n})}{s^2_{Y^0_n}(z_1,..,z_{2n})s_{Y_{2n}}(z_1,..,z_{2n})}.
\end{align}  

\section{Expectation value of \texorpdfstring{$\sigma_m^z$}{Lg}}
Frist, we need to renormalize our $R$-matrix (\ref{R}) in order to avoid singularities during 
the computations. We divide all the weights $a$, $b$ and $c$ by the weight $a$ and by abuse 
of notation denote the new weights by the same letters $a$, $b$ and $c$.

The operator $\sigma_m^z$ can be written in the F-basis \cite{MSdS,KMT}:
\begin{align}\label{Fsig}
\sigma_m^z=\prod_{i<m}\mathcal{T}(z_i) (A(z_m)-\kappa D(z_m)) \prod_{i>m}\mathcal{T}(z_i),
\end{align}
Now we sandwich this expression with the left and the right Bethe states. The result we 
can write as:
\begin{align}\label{sigexp}
\langle \sigma_m^z \rangle=2 \prod_{i=1}^{m-1}\tau_n(z_i)\prod_{i=m+1}^{n}\tau_n(z_i)
\langle 0|\prod_{i=1}^{n}C(\mu_i)A(z_m)\prod_{i=1}^{n}B(\zeta_i)|0\rangle-
\kappa S_n(\{\mu\};\{\zeta\}).
\end{align}
Using the Yang-Baxter algebra given by the RTT relations (\ref{RTT})
we can commute $A(z_m)$ in the first term through the $B$-operators. 
Noticing also that the prefactor of the first 
term in (\ref{sigexp}) is equal to $2\kappa/\tau_n(z_m)$ due to the property comming from the 
Bethe equations:
\begin{align}\label{beqprop}
\prod_{i=1}^{N}\tau_n(z_i)=\kappa,
\end{align}
we can write the expectation value $\sigma_m^z$ in the form:
\begin{align}\label{sigexp2}
\langle \sigma_m^z \rangle=S_n(\{\mu\};\{\zeta\})-
\sum_{a=1}^n f_{a} S_n(\{\mu\};\zeta_1,..,\hat{\zeta}_a,..,\zeta_n,z_m),
\end{align}
where $f_a$ are some coefficients depending on the Bethe roots 
(here, both $\mu_j$ and $\zeta_j$ are the Bethe roots). We omitted the overall factor of 
$\kappa$ in the last equation. More explicitly, the equation (\ref{sigexp2}) reads:
\begin{align}\label{sigexp3}
\langle \sigma_m^z \rangle= S_n(\zeta_1,..,\zeta_n)
-2\prod_{i=1}^n\frac{b(\zeta_i,z_m)}{a(\zeta_i,z_m)}
\sum_{i=1}^n \frac{c(\zeta_i,z_m)}{b(\zeta_i,z_m)}\prod_{j\neq i}\frac{a(\zeta_j,\zeta_i)}{b(\zeta_j,\zeta_i)}
S_n(z_m,\zeta_1,..,\hat{\zeta_i},..,\zeta_n).
\end{align}
Substituting the weights $a$, $b$ and $c$, the scalar product as in (\ref{spSL3}) 
and performing some algebraic manipulations, we obtain:
\begin{align}
&\langle \sigma_m^z \rangle= \tilde{S}_n\bigg{(}1+
6q\frac{Q(z_m)Q(q^2 z_m)}{F(q z_m)}G_n\bigg{)},\label{sigexp4}\\ 
&G_n=\sum_{i=1}^n \frac{\zeta_i z_m F(q\zeta_i)}{(\zeta_i-z_m)(q\zeta_i-z_m)
(\zeta_i-q z_m)Q(q^2\zeta_i)\prod_{j\neq i}(\zeta_i-\zeta_j)}. \label{Gfac}
\end{align}
The difference between (\ref{sigexp3}) and (\ref{sigexp4}) is that one of the two Vandermonde 
determinants in the denominator of $S_n$ is cancelled in (\ref{sigexp4}), thus we write 
$\tilde{S}_n$ in (\ref{sigexp4}). The rational function (\ref{Gfac}) is symmetric 
in the Bethe roots. We don't know how to write it compactly in terms of the symmetric 
functions of the Bethe roots.

Both, (\ref{sigexp3}) and (\ref{sigexp4}) can be rewritten 
back in the determinant form of a single matrix. Indeed, since the matrix elements of 
the Slavnov matrix have the form: 
\begin{align}\label{cmuk}
c_{i,k}=c_i(\mu_k),
\end{align}
what is written, e.g. in (\ref{sigexp3}) is the determinant of   
\begin{align}\label{cmuk2}
c_i(\mu_k)+\alpha_k c_i(z_m),
\end{align}
where $\alpha_k$ are the coefficients in the summation in (\ref{sigexp3}). 

Now let us set $q^3=1$. Using the expression for the scalar product at $q^3=1$ (\ref{spscur}) 
we arrive at:
\begin{align}\label{sigz1}
\langle \sigma_m^z \rangle = \text{const} \frac{1}{\prod_{i< j}(\zeta_j-\zeta_i)}
\det_{1\leq j,k\leq n}
\bigg{(} \frac{\zeta_k^{3j-2}}{F(q \zeta_k)} 
-2 \frac{z_m^{3j-2}Q(q \zeta_k)}{F(q z_m)Q(q z_m)}\bigg{)}.
\end{align}
where const depends on $q$ and thus can be omitted for simplicity.
 
Now we can turn to the final point of our work. We must symmetrize (\ref{sigz1}) with 
respect to the Bethe roots $\zeta_i$ 
using one or another symmetrization procedure. It is not an easy task and, we believe, there 
is a better way to do it than the one we choose here. Let us first write the eq.(\ref{sigz1}) as 
follows:
\begin{align}\label{sigz2}
\langle \sigma_m^z \rangle \propto \frac{1}{\prod_{i< j}(\zeta_j-\zeta_i)}
\det_{1\leq j,k\leq n}
\bigg{(} \zeta_k^{3j-2}F(q z_m)Q(q z_m) 
-2 z_m^{3j-2}F(q \zeta_k)Q(q \zeta_k)\bigg{)}.
\end{align}
This is a nicer, more symmetric, form of the determinant that we need to compute. It is 
also a determinant of a matrix with polynomial in $\zeta_k$ entries. 

Let us take again the symmetrization transform $\rho$ and apply it to 
a vector $(\zeta_1^l,..,\zeta_n^l)$. Using the properties (\ref{hspprop}) of the 
complete homogenious symmetric function we obtain:
\begin{align}\label{rhozetal}
\sum_{j=1}^n\rho_{i,j} \zeta_j^l=h_{i+l-n}.
\end{align}
Applying $\rho$ to the first term in (\ref{sigz2}) we get:
\begin{align}\label{rhozeta3j}
\sum_{k=1}^n\rho_{i,k} \zeta_k^{3j-2}=h_{i+3j-2-n}.
\end{align}
The second term we expand in $\zeta_k$:
\begin{align}\label{second1}
F(q\zeta_k)Q(q\zeta_k)=\sum_{j=0}^{3n}(-q)^j\zeta_k^{3n-j}\gamma_j,
\end{align}
where $\gamma_j$ are coefficients which depend only on the inhomogeneities and 
can be written as determinants. Their explicit form is given in the the appendix (\ref{gam}).
Applying $\rho$ to this term we get:
\begin{align}\label{second2}
\sum_{k=1}^n\rho_{i,k}F(q\zeta_k)Q(q\zeta_k)=\sum_{j=0}^{3n}(-q)^jh_{i+2n-j}\gamma_j.
\end{align}
Therefore, applying the transformation $\rho$ to the matrix elements in (\ref{sigz2}) 
gives us the final formula 
\begin{align}\label{sigz3}
\langle \sigma_m^z \rangle \propto \det_{1\leq j,k\leq n}
\bigg{(} h_{j+3k-2-n}F(q z_m)Q(q z_m) 
-2 z_m^{3k-2}\sum_{i=0}^{3n}(-q)^i h_{2n+j-i}\gamma_i\bigg{)}.
\end{align}
This formula still depends on the Bethe roots. However, each matrix element is a 
combination of homogenious symmetric polynomials of the Bethe roots. These and other 
symmetric functions of the Bethe roots, in fact, can be written as symmetric 
functions in the inhomogeneities. 
In the appendix we explain how to do that. In particular,
\begin{align}\label{fh}
h_i(\zeta_1,..,\zeta_n)=f_i(z_1,..,z_{2n}),
\end{align}
where $f_i$ is given in (\ref{felem}).
We believe that the expression (\ref{sigz3}) can be written in a much more simpler form. 
However, at this stage we don't know how to simplify it. This requires probably a 
different symmetrization of the eq.(\ref{sigz2}).

\section{Discussions}
We considered a particular interaction point $\Delta=-1/2$ of the ground state 
of the Heisenberg XXZ 
spin-$1/2$ chain with a twist $\kappa=q^2$. Our aim was to approach the correlation 
functions for the ground state of the model starting from the notion of the scalar products. 
The ground state scalar product turns out to be a simple Schur function (\ref{spscur}) 
which is a fully factorized polynomial (\ref{Aschur2}).
Using this result, in principle, we can compute the form factors and simple correlation 
functions corresponding to the ground state. In the last chapter we made an attempt to 
compute the expectation value of the $\sigma^z$ operators. Although, we arrived at a closed 
form expression (\ref{sigz3}) it still does not look as nice as we would like. 
Despite the simplicity of the scalar product it is still no easy to obtain 
a good formula for the simplest correlation functions.
We believe that the expression (\ref{sigz3}) can be further simplified.
One, perhaps, needs to use a different symmetrization method to deal with the 
determinant (\ref{sigz2}). 

Note that, our computations are restricted to the systems of even length. We expect that 
at odd lengths the same computation can be done. It is also an important problem 
to compute the correlation functions for the ground state of odd systems. 
Also, one could consider expectation values of operators 
with spin=$\pm 1/2$, e.g. $\sigma^+$ or $\sigma^-$. These expectation values are expected 
to be much easier than $\langle \sigma^z \rangle$.
An important result of our work is the integration of the Slavnov scalar products 
(\ref{spSL3}), which gives a new formula (valid for generic values of $q$). 
We started from the Slavnov 
respresentation of the scalar product, which is given by a matrix of derivatives 
of a fixed eigenvalue of the transfer matrix. Then applied a transformation which turned 
each entry of the matrix into a sum. The summation of the matrix elements yields 
back the transfer matrix eigenvalue plus a simple term.
 
\section*{Acknowledgements}
The author is grateful to Michael Wheeler and Paul Zinn-Justin for inspiring discussions. The 
work is supported by the ERC grant 278124 ``Loop models, Integrability and Combinatorics''.

\section*{Appendix: Symmetric functions of the Bethe roots}
Here we derive a number of identities satisfied by symmetric functions of the 
Bethe roots and symmetric functions of the spectral parameters. As a result 
we can express any symmetric polynomial that depends on the Bethe roots in terms of 
symmetric polynomials of spectral parameters. Let us introduce some notation:
\begin{align*}
&e_k^B=e_k(\zeta_1,..,\zeta_n),~~~h_k^B=h_k(\zeta_1,..,\zeta_n),~~~\\
&e_k^s=e_k(z_1,..,z_{2n}),~~~h_k^s=h_k(z_1,..,z_{2n}),
\end{align*}
where, as before, $\zeta$'s are the Bethe roots and $z$'s are the spectral parameters.  
We will derive this correspondence for both $e^B$ and $h^B$. From this a Schur function 
of the Bethe roots can be written using the Jacobi-Trudi or the dual Jacobi-Trudi 
identities (\ref{JTdJT}).

Let us first formulate our general statement. It is based on the 
proof of the equality of the Jacobi-Trudi with the dual Jacobi-Trudi determinants 
given e.g. in \cite{MD}. Consider three families of parameters: 
$a_i$, $b_i$ and $c_i$, $i\in\mathbb{Z}$, which satisfy the equations:
\begin{align}
&\sum_{i=0}^k a_i b_{k-i} =c_k,~~~k=0,1,2,..\label{abc1} \\
&a_i=b_i=c_i=0,~~~\text{for}~i<0 \label{abc2}
\end{align}
Construct three matrices:
\begin{align}\label{ABC}
A=(a_{i-j})_{0\leq i,j \leq n},~~~B=(b_{i-j})_{0\leq i,j \leq n},~~~\text{and}~~
C=(c_{i-j})_{0\leq i,j \leq n}.
\end{align}
These matrices are lower triangular due to (\ref{abc2}).
The equations (\ref{abc1}) are equivalent to the matrix equation:
\begin{align}\label{ABeqC}
A B =C,
\end{align}
or, if we define a lower triangular matrix $D=C^{-1}A$, $D=(d_{i-j})_{0\leq i,j \leq n}$ 
with entries 
\begin{align}\label{delem}
d_k=(-1)^k\det \left(
\begin{array}{cccc}
 \frac{a_0}{c_0} & \frac{a_1}{c_0} & \dots & \frac{a_k}{c_0} \\
 1 & \frac{c_1}{c_0} & \dots & \frac{c_k}{c_0} \\
 \vdots & \vdots & \vdots  & \vdots \\
 0 & 0 &  \dots & \frac{c_1}{c_0} \\
\end{array}
\right),
\end{align}
then we can write
\begin{align}\label{DB}
D B =I.
\end{align}
This means that  the matrix $D$ is the inverse of the matrix $B$.  
It follows that each minor of $D$ is equal to the complementary cofactor of 
the transposed matrix $B$. Recall that for a partition $\nu=(\nu_1,..,\nu_s)$ with 
$\nu_1= p$, the $p+s$ numbers: 
\begin{align}\label{nups}
\nu_i+s-i~~~(1\leq i\leq s),~~~~s-1+i-\nu_i^{\prime}~~~(1\leq i\leq p),
\end{align}
form a permutation of $(0,1,..,p+s-1)$. The prime of $\nu$ is the transposed partition 
of $\nu$. Consider the minor of $D$ with row indices $\nu_i+s-i$ ($1\leq i\leq s$) 
and column indices $s-i$ ($1\leq i\leq s$). This minor is equal up to a sign to the 
complementary cofactor of the 
transposed $B$ which has row indices $s-1+i-\nu_i^{\prime}$ ($1\leq i\leq p$) and column 
indices $s-1+i$ ($1\leq i\leq p$). We obtain
\begin{align}\label{deqb}
\det_{1\leq i,j \leq s}d_{\nu_i-i+j}=(-1)^{\sum_{i=1}^s\nu_i}
\det_{1\leq i,j \leq s}b_{\nu_i^{\prime}-i+j}.
\end{align}

Let us look at some application of this formula. If we set $b_i=(-1)^i e_i$ and 
$a_i= h_i$, where $e_i$ and $h_i$ are the elementary symmetric polynomials and the 
homogeneous symmetric polynomials, respectively, then the conditions (\ref{abc1}) 
and (\ref{abc2}) are satisfied with the matrix $C$ being the identity matrix. Taking this 
into account in (\ref{delem}), we get $d_k=h_k$, and (\ref{deqb}) reads
\begin{align}\label{JTdJT}
s_{\nu}=\det_{1\leq i,j \leq s}h_{\nu_i-i+j}=
\det_{1\leq i,j \leq s}e_{\nu_i^{\prime}-i+j},
\end{align}
This is the Jacobi-Trudi and the dual Jacobi-Trudi identities for the Schur function 
$s_{\nu}$.

Now we will use the $T-Q$ realtion (\ref{TQ}) do derive equations of the form 
(\ref{abc1}) for various symmetric polynomials.
Let us first recall the expression from the Appendix of the paper \cite{PDFPZJ}
\begin{align}\label{FQdet}
F(q^2 t)Q(q^2 t)=
\frac{
\left(
\begin{array}{ccccccc}
 z_1^{3n} & z_2^{3n} & \dots & z_{2n}^{3n} &  t^{3n} \\
 \vdots   &   \vdots &       & \vdots      & \vdots        \\
 z_1^{3k+2} & z_2^{3k+2} & \dots & z_{3n}^{3k+2} &  t^{3k+2} \\
 z_1^{3k} & z_2^{3k} & \dots & z_{3n}^{3k} &  t^{3k} \\
 \vdots   &   \vdots &       & \vdots      & \vdots        \\
  z_1^2 & z_2^2 & \dots & z_{3n}^2 &  t^2 \\
 1 & 1 & 1 & \dots & 1 \\
\end{array}
\right)}
{\left(
\begin{array}{ccccccc}
 z_1^{3n-1} & z_2^{3n-1} & \dots & z_{2n}^{3n-1} \\
 \vdots   &   \vdots &       & \vdots              \\
 z_1^{3k+2} & z_2^{3k+2} & \dots & z_{3n}^{3k+2}  \\
 z_1^{3k} & z_2^{3k} & \dots & z_{3n}^{3k}  \\
 \vdots   &   \vdots &       & \vdots              \\
  z_1^2 & z_2^2 & \dots  \\
 1 & 1 & \dots & 1  \\
\end{array}
\right)
},
\end{align}
with $Y_{2n}^{\prime}=\{n,n,..,1,1,0\}$.
This expression gives the $Q$ function and hence all coefficients of the expansion 
of the $Q(q^2 t)$ in powers of $t$. These coefficients are the elementary symmetric polynomial 
of the Bethe roots:
\begin{align}\label{Qesp}
Q(t)=q^{2n} \sum_{i=0}^{n}(-1)^i t^{n-i} q^{i}e^B_{i}.
\end{align}
We would like to redefine the $e^B$ in order to avoid carrying around the factors of $q$. Let 
$\tilde{e}_i^B=q^i e^B_i$, and we omit the tilde in this appendix by abuse of notation.
     The function $F$ in (\ref{FQdet}) 
is the generating function of the elementary symmetric polynomials in the spectral 
parameters:
\begin{align}\label{Fesp2}
F(t)=q^{4n}\sum_{i=0}^{2n}(-1)^i t^{2n-i} e_{i}^s.
\end{align}
Now we expand both sides of the equation (\ref{FQdet})
in powers of $t$:
\begin{align}\label{QFexp}
\sum_{k=0}^{3n}(-1)^k t^{3n-k} \sum_{j=0}^{k}e^B_{j} e^s_{k-j}=
\sum_{k=0}^{3n}(-1)^k t^{3n-k} \gamma_{k}.
\end{align}
The coefficients $\gamma_k$ are Schur functions which are labeled by partitions $\pi_k$ 
divided by the Schur function $s_{\tilde{Y}_n}$. 
These Schur functions can be obtained from the expansion in $t$ of the Schur 
function $s_{Y^{\prime}_{2n+1}}$ in (\ref{FQdet}). The partitions $\pi_k$ are derived from 
$Y^{\prime}$ in the following way.
The partition $Y^{\prime}_{2n+1}$ has length equal to $2n+1$, and we can derive $2n+1$ other 
partitions as follows:
\begin{align}\label{parti2}
&U^{(j)}=Y^{\prime} + \theta^{(j)},\\
&\theta^{(j)}_{i} = \left\{
  \begin{array}{l l}
    0 & \quad \text{if $i\geq j$}\\
    1 & \quad \text{otherwise.}
  \end{array} \right. \nonumber
\end{align}
The partitions $\pi_k$ are written in terms of $U^{(k)}$ as:
\begin{align}\label{parti3}
\pi_{k-1}=\{U_1^{(k)},..,\hat{U}_k^{(k)},..,U_{2n+1}^{(k)}\},
\end{align}
so $\pi_k$ is the $U^{(k)}$ with the $k+1$'st part absent ($k=0,..,2n$). 
The expansion of the 
Schur function $s^{\prime}_{Y_{2n+1}}$ in (\ref{FQdet}) gives $3n+1$ coefficients $\gamma_i$ 
some of which are equal to zero, others are Schur functions of $\pi_k$: 
\begin{align}\label{gam}
&\gamma_{3j}=(-1)^{j}\frac{s_{\pi_{2j}}}{s_{\tilde{Y}}},~~~~j=0,..,n\nonumber \\
&\gamma_{3j+1}=(-1)^{j}\frac{s_{\pi_{2j+1}}}{s_{\tilde{Y}}},~~~~\gamma_{3j+2}=0,~~~~j=0,..,n-1.
\end{align}

Let us get back to (\ref{QFexp}). Collecting the coefficients over $t$ we get:
\begin{align}\label{eg}
\sum_{j=0}^{k}e^B_{j} e^s_{k-j}=\gamma_{k}.
\end{align}
Setting $a$ to $e^s$, $b$ to $e^B$ and $c$ to $\gamma$ (note that $\gamma_0=1$), 
and write in this particular case $f$ instead of $d$, we obtain:
\begin{align}\label{feqeB}
\det_{1\leq i,j \leq s}f_{\nu_i-i+j}=
\det_{1\leq i,j \leq s}e^B_{\nu_i^{\prime}-i+j}.
\end{align}
with 
\begin{align}\label{felem}
f_k=(-1)^k\det \left(
\begin{array}{cccc}
 1 & e^s_1 & \dots & e^s_k \\
 1 & \gamma_1 & \dots & \gamma_k \\
 \vdots & \vdots & \vdots  & \vdots \\
 0 & 0 &  \dots & \gamma_1 \\
\end{array}
\right),
\end{align}
which effectively means that $h_k^B=f_k$. The polynomials $f_k$ can also be written as
a sum:
\begin{align}\label{fsum}
f_k=\sum_{j=0}^k (-1)^j e_{k-j}\delta_j,
\end{align}
where $\delta_j$ are the entries of the inverse of $C$ with $c_i=\gamma_i$ which can 
be written as
\begin{align}\label{delta}
\delta_k=\det_{1\leq i,j\leq k} \gamma_{1-i+j}.
\end{align}
We also can write $\gamma_k$ in terms of $\delta_i$: 
\begin{align}\label{gamelem}
\gamma_k=\det_{1\leq i,j\leq k} \delta_{1-i+j}.
\end{align}
Since $\gamma$ and $\delta$ are dual we can write the equations (\ref{abc1}) for them:
\begin{align}\label{gamdelt}
\sum_{i=0}^k (-1)^i\gamma_i \delta_{k-i} =0,~~~k=0,1..
\end{align}
Now if we take a sum of $\gamma$ with the $f$'s the previous equation together 
with the eq.(\ref{fsum}) allow to obtain the polynomials $e^s$:
\begin{align}\label{gamf}
\sum_{i=0}^k \gamma_i f_{k-i} =e^s_k,~~~k=0,1..,2n.
\end{align}
Since $f_i=h^B_i$, we get another equation for a determinat of the Bethe roots:
\begin{align}\label{geqhB}
\det_{1\leq i,j \leq s}g_{\nu^{\prime}_i-i+j}=(-1)^{\sum_j \nu_j}
\det_{1\leq i,j \leq s}h^B_{\nu_i-i+j},
\end{align}
where $g_i$ can be written as a determinant:
\begin{align}\label{gelem}
g_k=\det \left(
\begin{array}{cccc}
 1 & \gamma^s_1 & \dots & \gamma^s_k \\
 1 & e^s_1 & \dots & e^s_k \\
 \vdots & \vdots & \vdots  & \vdots \\
 0 & 0 &  \dots & e^s_1 \\
\end{array}
\right),
\end{align}
and also as a sum
\begin{align}\label{gsum}
g_k=\sum_{j=0}^k (-1)^k h^s_{k-j}\gamma_j.
\end{align}
It follows also that $e^B_i=g_i$.
Let us summarize what we obtained:
\begin{align}
&e^s_k=\sum_{i=0}^k \gamma_{k-i}h_i^B,
~~~~~~~~~h^s_k=\sum_{i=0}^k (-1)^i\delta_{k-i}e_i^B,\label{Btos}\\
&e^B_k=\sum_{i=0}^k \gamma_{k-i}h_i^s,
~~~~~~~~~h^B_k=\sum_{i=0}^k (-1)^{i+k}\delta_{k-i}e_i^s.\label{stoB}
\end{align}

\small{\bibliographystyle{plain}}
\bibliography{bibSP}

\end{document}